\documentclass[prb,twocolumn,preprintnumbers,amsmath,amssymb]{revtex4}

\usepackage{graphicx}
\usepackage{dcolumn}
\usepackage{bm}
\usepackage{times}
\usepackage{mathptmx}
\usepackage{color}

\begin{document}

\title{Mutual induction of magnetic $3d$ and $4f$ order\\
in multiferroic hexagonal ErMnO$_3$}

\author{D. Meier,$^{1,2,\ast}$ H. Ryll,$^3$ K. Kiefer,$^3$ B. Klemke,$^3$ J.-U. Hoffmann,$^3$ R. Ramesh,$^{1,2,4}$ and M. Fiebig$^5$}

 \affiliation{$^1$Department of Materials Science and Engineering, University of California, Berkeley, CA 94720, USA}
 \affiliation{$^2$Department of Physics, University of California, Berkeley, CA 94720, USA}
 \affiliation{$^3$Helmholtz-Zentrum Berlin f\"{u}r Materialien und Energie, 14109 Berlin, Germany}
 \affiliation{$^4$Materials Science Devision, Lawrence Berkeley National Laboratory, Berkeley, CA 94720, USA}
 \affiliation{$^5$Department of Materials, ETH Zurich, 8093 Zurich, Switzerland}
\date{\today}

\begin{abstract}
The complex interplay between the $3d$ and $4f$ moments in hexagonal ErMnO$_3$ is investigated by
magnetization, optical second harmonic generation, and neutron-diffraction measurements. We revise
the phase diagram and provide a microscopic model for the emergent spin structures with a special
focus on the intermediary phase transitions. Our measurements reveal that the $3d$ exchange
between Mn$^{3+}$ ions dominates the magnetic symmetry at $10\text{~K}<T<T_{\rm N}$
with Mn$^{3+}$ order according to the $\Gamma_4$ representation triggering $4f$ ordering according
to the same representation on the Er$^{3+}$(4b) site. Below 10~K the magnetic order is governed by
$4f$ exchange interactions of Er$^{3+}$ ions on the 2a site. The magnetic Er$^{3+}(2a)$ order
according to the representation $\Gamma_2$ induces a magnetic reorientation
($\Gamma_4\to\Gamma_2$) at the Er$^{3+}(4b)$ and the Mn$^{3+}$ sites. Our findings highlight the
fundamentally different roles the Mn$^{3+}$, $R^{3+}$(2a), and $R^{3+}$(4b) magnetism play in
establishing the magnetic phase diagram of the hexagonal \textit{R}MnO$_3$ system.
\end{abstract}

\maketitle

\section{Introduction: multiferroic hexagonal manganites}

Materials with coexisting magnetic and electric order, the so-called multiferroics, have been
attracting a lot of attention since it was recognized that they can display gigantic
magnetoelectric coupling effects: Magnetic properties can be controlled by electric fields and
vice versa.\cite{Kimura03a,Hur04a,Fiebig05a,Cheong07a} In this context the hexagonal (h-)
manganites h-$R$MnO$_3$ with $R$ = Sc, Y, Dy--Lu play an exceptional role  because they offer
great flexibility for tuning such magnetoelectric correlations and studying the coupling between
spin, charge, and lattice degrees of freedom in multiferroics. Crystals of the h-$R$MnO$_3$ family
are structurally equivalent and display ferroelectric order below about 1000~K with a spontaneous
polarization of $\approx 5.6$~$\mu$C/cm$^2$ along the hexagonal
$c$-axis.\cite{Coeure66a,Pauthenet70a,Kumagai12a} The $R^{3+}$ ions vary in size and magnetic
moment and due to the interaction with the Mn$^{3+}$ ions the variety of magnetic phases and
magnetoelectric interaction phenomena emerging below about 100~K is particularly rich. This
includes contributions to the magnetization induced by ferroelectric poling, giant magneto-elastic
coupling effects, and a coupling between magnetic and ferroelectric domain
walls.\cite{Lottermoser04a,Lee08a,Fiebig02b}

A discussion of the magnetoelectric coupling phenomena in the h-$R$MnO$_3$ compounds with $R$ =
Dy--Yb invariably involves the magnetic $4f$ moments. Yet, investigations of the rare-earth order
had remained scarce for a long time. Only recently, earlier magnetization measurements have been
complemented by a structural analysis of the $R^{3+}$ order in h-HoMnO$_3$, h-YbMnO$_3$, and
h-DyMnO$_3$ by neutron or resonant x-ray diffraction. The studies revealed that the $3d$--$4f$
interaction in h-\textit{R}MnO$_3$ is more complex than previously
assumed\cite{Fabreges08a,Harikrishnan09a,Lonkai02a,Sugie02a,Munawar06a,Tyson10a} and that the
magnetic Mn$^{3+}$ and $R^{3+}$ lattices can have a different space
symmetry.\cite{Wehrenfennig10a} Yet, with very few exceptions\cite{Fabreges08a} little is known about the $4f$--$4f$ exchange interaction
between $R^{3+}$ moments occupying \textit{different} Wyckoff positions, i.e., the 4b and 2a sites
of the hexagonal unit cell (see inset of Fig.~\ref{fig:fig3} (a) for a schematic
illustration). A detailed knowledge, however, is indispensable for understanding
the complex magnetic, multiferroic, and magnetoelectric coupling processes in h-$R$MnO$_3$. It
becomes even more important in view of the current intensive studies addressing the domain
walls\cite{Meier12a,Meier12b,Wu12a,Du11a} and their magnetic properties.\cite{Geng12a} Here, any
statement about the \textit{local} magnetic properties initially requires a precise knowledge of
the \textit{global} bulk spin structure. Consequently, the first goal is to develop a model
explaining the magnetic phase diagram of the h-$R$MnO$_3$ series in general.

For our study we have chosen h-ErMnO$_3$ --- its magnetic Mn$^{3+}$ phase diagram is similar to
that of h-TmMnO$_3$ and h-YbMnO$_3$ which establishes it as a prototypical compound within the
h-$R$MnO$_3$ series. Based on magnetoelectric and magnetization measurements a magnetic Er$^{3+}$
order was proposed\cite{Iwata98a,Sugie02a} but not verified or uniquely related to the coexisting
Mn$^{3+}$ order.

In this Report we clarify the spin structure of the Mn$^{3+}$ and Er$^{3+}$ sublattices and
introduce a microscopic model coherently explaining the magnetic phase diagram of h-ErMnO$_3$ that
can be projected onto the h-$R$MnO$_3$ system in general. For this purpose we performed
complementary magnetization, second harmonic generation (SHG), and neutron-diffraction
measurements at temperatures down to 30~mK. Below the N\'{e}el temperature we find
antiferromagnetic order of the Mn$^{3+}$ moments triggering the magnetic order of the
Er$^{3+}$(4b) site according to the same representation ($\Gamma_4^{\rm{Er,Mn}}$) while the
Er$^{3+}$(2a) sites remain disordered. In contrast, the ground state toward 0~K is ferrimagnetic
with \textit{all} spins ordered according to the same representation ($\Gamma_2^{\rm{Er,Mn}}$).
The intermediary phase transition ($\Gamma_4^{\rm{Er,Mn}}\to\Gamma_2^{\rm{Er,Mn}}$) occurs via a
transient breakdown of the magnetic order on the Mn$^{3+}$ and Er$^{3+}$(4b) sites so that the
highest possible symmetry is maintained. The study advances our understanding of the different
roles the Mn$^{3+}$, $R^{3+}$(2a), and $R^{3+}$(4b) moments play in establishing the magnetic
phase diagram of the h-\textit{R}MnO$_3$ system.

\section{Experimental results}

\subsection{Temperature- and field-dependent magnetization measurements} \label{magnetization}

In order to clarify the spin structure in h-ErMnO$_3$ we scrutinized the magnetic phase diagram
and extended it towards the milli-Kelvin regime to capture the actual ground state. Our
magnetization measurements were performed at the LaMMB-MagLab of the Helmholtz-Zentrum Berlin
using a h-ErMnO$_3$ single crystal of 20~mg grown by the floating-zone technique. For measurements
above 1.8~K a standard vibrating sample magnetometer was used, whereas cantilever magnetometry was
applied to study magnetic transitions below 1.8~K.

\begin{figure}[h]
\centering
\includegraphics[width=8.5cm,keepaspectratio,clip]{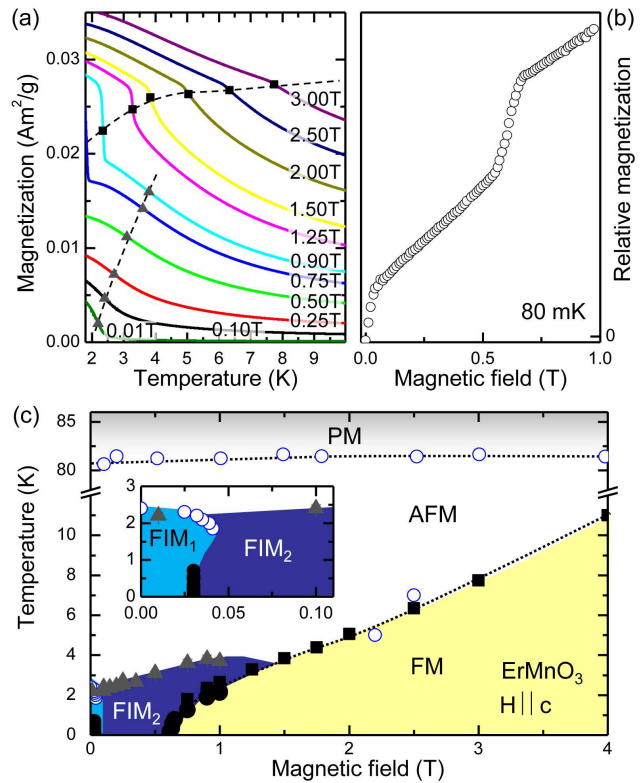}
\caption{\label{fig:fig1} (a) Temperature dependence of the magnetization $M(T)$ in h-ErMnO$_3$
measured after field cooling. The data set, recorded with $dT/dt=+0.2$~K/min in magnetic fields
applied parallel to the $c$-axis, reveals two different anomalies indicated by gray triangles and
black squares, respectively. As guide to the eye dashed lines indicate the magnetic-field
dependence of the two transitions (b) Magnetic-field dependence of the magnetization $M(H)$ at
80~mK. The curve shows two transistions that manifest as a step-like increase in $M(H)$. (c)
HT-phase diagram of h-ErMnO$_3$. Triangles and squares correspond to transitions deduced form
$M(T)$ (see (a)) while filled circles represent transitions observed in $M(H)$. Blue open circles
are additional data points taken from Ref.~\onlinecite{Yen07a}.}
\end{figure}

Figure~\ref{fig:fig1}(a) shows magnetization data for $M(T)$ taken with increasing temperature
$dT/dt=+0.2$~K/min after field cooling ($H\,\|\,c$). Note that identical curves were obtained for
$dT/dt<0$ (not shown) indicating non-hysteretic behavior. For the remainder of this work we
therefore restrict the discussion to measurements with $dT/dt>0$.

As seen in Fig.~\ref{fig:fig1}(a), at 0.01~T a magnetic moment of $4\cdot 10^{-3}$~Am$^2$/g
is measured at 1.8~K. It decreases rapidly with increasing temperature showing a
minimum in its derivative $dM/dT$ around 2.2~K. The result is in agreement with earlier
publications where the magnetization was attributed to long range magnetic order in the Er$^{3+}$
sublattice.\cite{Sugie02a} For increasing magnetic fields the associated change in $M(T)$ becomes
less pronounced and the minimum in $dM/dT$ shifts to higher temperature (see gray triangles in
Fig.~\ref{fig:fig1}(a)). The same trend is observed in magnetic torque measurements revealing a
pronounced change in response between 1~K and 4~K (not shown). A second anomaly manifests at
0.75~T as a step-like increase in $M(T)$. This anomaly shifts continuously to higher temperatures
when the magnetic field is increased further (see black circles in Fig.~\ref{fig:fig1}(a)).
Figure~\ref{fig:fig1}(c) summarizes the results by showing the magnetic phase diagram in the
temperature/magnetic-field plane. In order to trace the phase boundaries we investigated the
magnetic-field dependence of the magnetization $M(H)$ down to the milli-Kelvin range as
exemplified by Fig.~\ref{fig:fig1}(b). All low-temperature magnetic-field scans clearly indicate
two consecutive anomalies around 0.03~T and 0.6~T, respectively. Hence, based on our magnetization
measurements we have to distinguish four different magnetically ordered states which are denoted
as AFM, FIM$_1$, FIM$_2$, and FM in the phase diagram in Fig.~\ref{fig:fig1}(c).

The detailed analysis of the nature of these states, i.e.\ the corresponding spin structures and
interactions, will be the topic of the following sections. We will distinguish four
representations (magnetic space symmetries): $\Gamma_1$ ($P6_3cm$), $\Gamma_2$
($P6_3\underline{cm}$), $\Gamma_3$ ($P\underline{6}_3c\underline{m}$), and $\Gamma_4$
($P\underline{6}_3\underline{c}m$). As we will see, lower symmetries like $P6_3$ or
$P\underline{6}_3$ do not have to be considered. For the sake of clarity we will first focus on
the magnetic order of the Mn$^{3+}$ lattice, followed by a study of the Er$^{3+}$ order at the 4b
and 2a sites. In the third part we will associate these spin structures to the magnetic phases
identified in Fig.~\ref{fig:fig1}(c). An investigation of the intermediary transitions connecting
the different magnetic phases follows. It will allow us to describe the sequence of magnetic phase
transitions in a comprehensive model based on the coupling between the magnetic sublattices in the
h-ErMnO$_3$ system with an outlook to the h-$R$MnO$_3$ system in general.

\subsection{SHG and neutron-diffraction study of the Mn$^{3+}$ spin structure} \label{Mn-order}

The magnetic Mn$^{3+}$ moments order antiferromagnetically at $T_{\rm N}\approx
80$~K.\cite{Koehler64a} The resulting structure has been discussed for more than five decades.
Based on the extensive literature on h-ErMnO$_3$ the analysis and discussion can already be
restricted to $\Gamma_2^{\rm{Mn}}$ and $\Gamma_4^{\rm{Mn}}$ as only possible magnetic
representations for the Mn$^{3+}$ order (see Fig.~\ref{fig:fig4} for a schematic
illustration).\cite{Fiebig02a,Sekhar05a,Park02a,Lee06a} In order to scrutinize the emergence of
the magnetic order according to either of these representations we performed complementary
experiments by optical SHG and by neutron diffraction on selected site-specific reflections. The
neutron-diffraction experiments were performed at the E2 beamline of the Helmholtz-Zentrum Berlin.
A h-ErMnO$_3$ single crystal of $2\times 3\times 5$~mm$^3$ (180~mg) was cut from the same batch as
the crystal studied by magnetometry. The sample was mounted in a $^3$He/$^4$He dilution insert and
investigated in the $h0l$ plane at a wavelength of {2.39~\AA}. Since neutron-diffraction
experiments alone prohibit a unique distinction between the two aforementioned space
groups,\cite{Park02a} SHG measurements were also conducted in order to uniquely determine the
magnetic Mn$^{3+}$ order. SHG is described by the equation
$P_i(2\omega)=\epsilon_0\chi_{ijk}E_j(\omega)E_k(\omega)$. A light field $\vec{E}$ at frequency
$\omega$ is incident onto a crystal, inducing a dipole oscillation $\vec{P}(2\omega)$, which acts
as source of a frequency-doubled light wave. The susceptibility $\chi_{ijk}$ couples incident
light fields with polarizations $j$ and $k$ to a SHG contribution with polarization $i$. The
magnetic and crystallographic symmetry of a compound is uniquely related to the set of nonzero
components $\chi_{ijk}$ and, therefore, allows to distinguish $\Gamma_2^{\text{Mn}}$ and
$\Gamma_4^{\text{Mn}}$.\cite{Fiebig05b} Note that SHG has been applied earlier for investigating
the magnetic structure of the Mn$^{3+}$ sublattice in h-ErMnO$_3$ but never before neutron
diffraction and SHG were applied to the \textit{same} sample and verified for the consistency of
the two techniques.

\begin{figure}[h]
\centering
\includegraphics[width=8.5cm,keepaspectratio,clip]{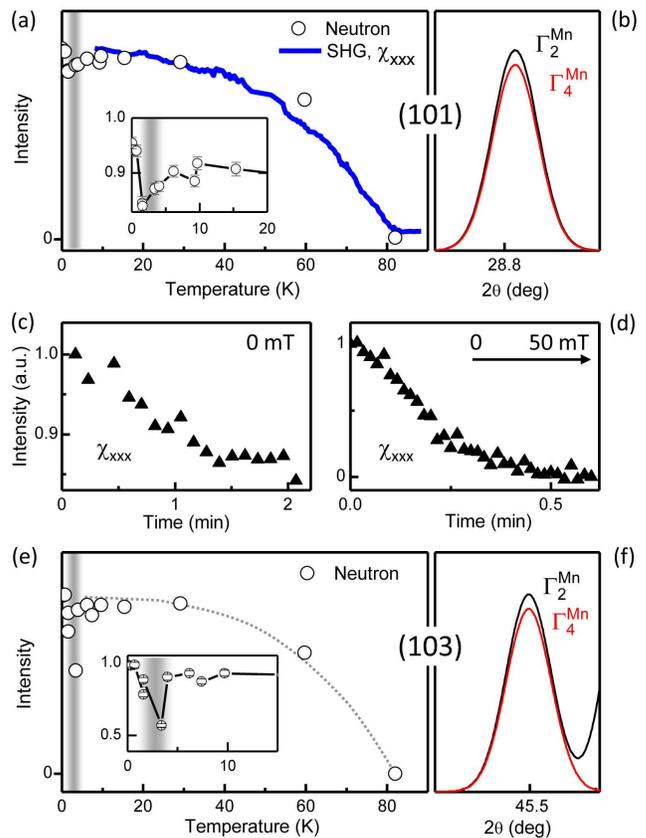}
\caption{\label{fig:fig2} (a) Comparison of optical SHG measurements and neutron-diffraction
experiments. Both the $\chi_{xxx}$ component of the nonlinear susceptibility tensor and the
magnetic (101) reflection arise at $T_{\rm N}=80$~K. Between 1~K and 4~K the (101) peak intensity
displays a dip which is shown in detail in the inset to Fig.~\ref{fig:fig2}(a). (b) SIMREF2.6
simulation of the relative (101) peak intensity assuming Mn$^{3+}$ order according to
$\Gamma_2^{\rm{Mn}}$ and $\Gamma_4^{\rm{Mn}}$. (c), (d) Time dependence of the SHG signal from
$\chi_{xxx}$ after decreasing the temperature from 5 to 1.8~K. The steady gradual decrease
observed in (c) is supported by a minuscule magnetic field $H\,\|\,c$ ramped from 0 to 0.05~T
linearly with time in (d). (e) Temperature dependence of the (103) reflection probing the
Mn$^{3+}$ order as corroborated by the simulations presented in (f). The gray dotted line retraces
the SHG data shown in (a) serving as guide to the eye. In agreement with (a) the (103) peak
intensity transiently breaks down between 1~K and 4~K.}
\end{figure}

Figure~\ref{fig:fig2}(a) shows a comparison of temperature-dependent neutron-diffraction and SHG
intensities, the former for the (101) reflection and the latter for the $\chi_{xxx}$ component of
the nonlinear susceptibility tensor.\cite{Fiebig02a,Park02a} Both signals arise below $T_{\rm N}$
and exhibit the same temperature dependence emphasizing that SHG is coupling to the
antiferromagnetic order of the Mn$^{3+}$ moments. The SHG measurement proves that the Mn$^{3+}$
moments of h-ErMnO$_3$ order according to the $\Gamma_4^{\text{Mn}}$ representation in the
temperature range $10~{\rm K}<T<T_{\rm N}$, clearly discarding the $\Gamma_2^{\text{Mn}}$ symmetry
on the basis of selection rules: Only $\Gamma_4^{\text{Mn}}$ allows SHG from $\chi_{xxx}$ whereas
$\Gamma_2^{\text{Mn}}$ does not.\cite{Fiebig05b,Kordel09a}

Below about 2~K the SHG intensity from $\chi_{xxx}$ begins to drop with time. According to
Fig.~\ref{fig:fig2}(c) a decrease by 20\% is obtained during the first 90~s after reducing the
temperature from 5 to 1.8~K. The decrease slows down afterwards. Note that this
behavior is reproducible and not caused by changes of the linear optical properties during
cooling. A minuscule magnetic field accelerates the decrease of the SHG signal.
Figure~\ref{fig:fig2}(d) shows that it steadily drops to zero within 30~s when a magnetic field is
ramped from 0 to 0.05~T linearly with time. The gradual decrease of the $\chi_{xxx}$ component
from its full value to zero is characteristic for a $\Gamma_4^{\text{Mn}}\to\Gamma_2^{\text{Mn}}$
transition of the Mn$^{3+}$ sublattice.\cite{Fiebig02a,Wehrenfennig10a,Fiebig00a,Fiebig05c} This
is contrasted by only a minor dip in the neutron diffraction at the (101) peak between 1~K and
4~K. The persistence of the (101) reflection is expected  as $\Gamma_4^{\text{Mn}}$ and
$\Gamma_2^{\text{Mn}}$ representations lead to (101) peak intensities differing by only 7\% (see also
Fig.~\ref{fig:fig2}(b)).\cite{Park02a,Lee06a} The dip emerging at the intermediary transition,
however, cannot be explained on the basis of symmetry.

We therefore verified its occurrence by repeating the temperature dependent measurement on a
second magnetic reflection coupling to the Mn$^{3+}$ order. For this purpose we chose the (103)
reflection as SIMREF2.6 simulations\cite{Ritter99a} (see Fig.~\ref{fig:fig2}(f)) reveal no
intermixing with Er$^{3+}$ contributions.\cite{footnote1} Figure~\ref{fig:fig2}(e) shows a dip
between 1~K and 4~K that is even more pronounced compared to the one seen in
Fig.~\ref{fig:fig2}(a). It points to a transient breakdown of the Mn$^{3+}$ order that will be
discussed in Section~\ref{model}. Note that the dip in the Mn$^{3+}$-related intensity was not
observed in earlier neutron measurements on the same sample where the base temperature of the
experiment was limited to 1.8~K.\cite{Tomuta03a} This emphasizes the importance of establishing a
well-defined ground state, here by entering the milli-Kelvin regime, for performing an accurate
analysis of the magnetic order in this strongly frustrated system.

\subsection{Neutron-diffraction study of the Er$^{3+}$ spin structure} \label{Er-order}

Analogous to the case of the Mn$^{3+}$ moments four different representations denoted
$\Gamma_1^{\text{Er}}$ to $\Gamma_4^{\text{Er}}$ have to be distinguished for the Er$^{3+}$
order.\cite{Fabreges08a,Munoz00a,Nandi08a,Fiebig03a} However, in contrast to the Mn$^{3+}$ lattice
no SHG contributions coupling to the rare-earth system were found so that our symmetry analysis is
entirely based on neutron data.

\begin{figure}[h]
\centering
\includegraphics[width=8.5cm,keepaspectratio,clip]{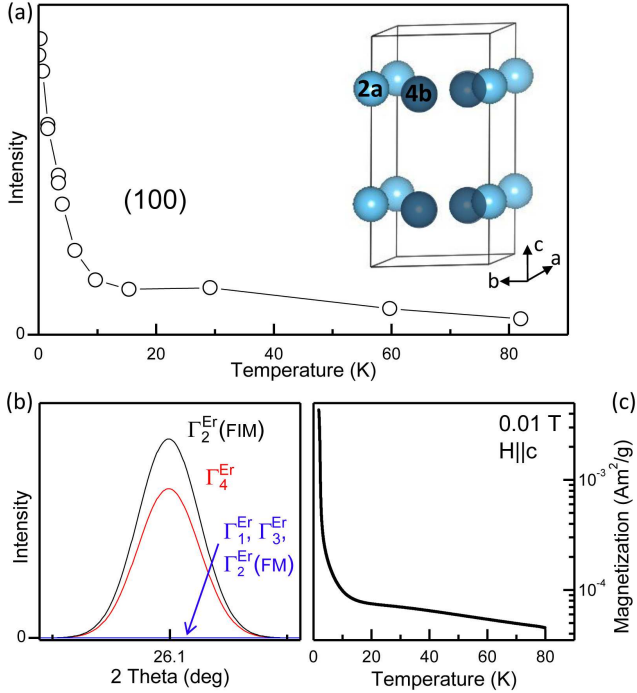}
\caption{\label{fig:fig3} (a) Temperature dependence of the (100) peak intensity and schematic
illustration of the Er$^{3+}$(2a) and Er$^{3+}$(4b) positions in the hexagonal unit cell of
h-ErMnO$_3$. (b) SIMREF2.6 simulation reveal  that a (100) reflection can emerge only for
Er$^{3+}$ order according to the $\Gamma_4^{\rm{Er}}$ or $\Gamma_2^{\rm{Er}}(\rm{FIM})$
representation with the latter one leading to the stronger reflection. (c) $M(T)$ measurement
taken at 10~mT exhibiting a similar temperature dependence as the (100) peak intensity.}
\end{figure}

In Fig.~\ref{fig:fig3}(a) we present the temperature dependence of the (100) reflection of
h-ErMnO$_3$. Since neither $\Gamma_4^{\rm{Mn}}$ order nor $\Gamma_2^{\rm{Mn}}$ order allows
Mn$^{3+}$ contributions to the (100) peak, any contribution to this peak has to be attributed to
the magnetic order of the Er$^{3+}$ moments.\cite{Wehrenfennig10a} The (100) peak emerges at 80~K.
Its intensity increases monotonously with decreasing temperature down to about 10~K where a change
of slope occurs. Thus, according to Fig.~\ref{fig:fig3}(a) the Er$^{3+}$ sublattice does not order
at 3~K as reported before,\cite{Sugie02a} but already at $T_{\rm N}$. Yet, the
temperature dependence of the Er$^{3+}$-related signal is different from those related to
Mn$^{3+}$ in Fig.~\ref{fig:fig2}. This is reminiscent of the situation in h-DyMnO$_3$ and
h-YbMnO$_3$ and indicates that the magnetic Er$^{3+}$ order is related to the antiferromagnetic
Mn$^{3+}$ order by the magnetic triggering mechanism identified in
Ref.~\onlinecite{Wehrenfennig10a}. The triggering mechanism describes a biquadratic
order-parameter coupling between two subsystems of a compound. Ordering in one of them can induce
ordering in the other at the same temperature whenever the associated coupling term lowers the
ground state energy.\cite{Holakowsky73} Because of the biquadratic nature of the coupling the
Er$^{3+}$ and Mn$^{3+}$ sublattices can order according to different magnetic space groups at
$T_{\rm N}$. Thus, in spite of the known magnetic order of the Mn$^{3+}$ system, the
magnetic Er$^{3+}$ system may order according to any of the four representations
$\Gamma_1^{\rm{Er}}$ to $\Gamma_4^{\rm{Er}}$.

\begin{figure}[h]
\centering
\includegraphics[width=8.5cm,keepaspectratio,clip]{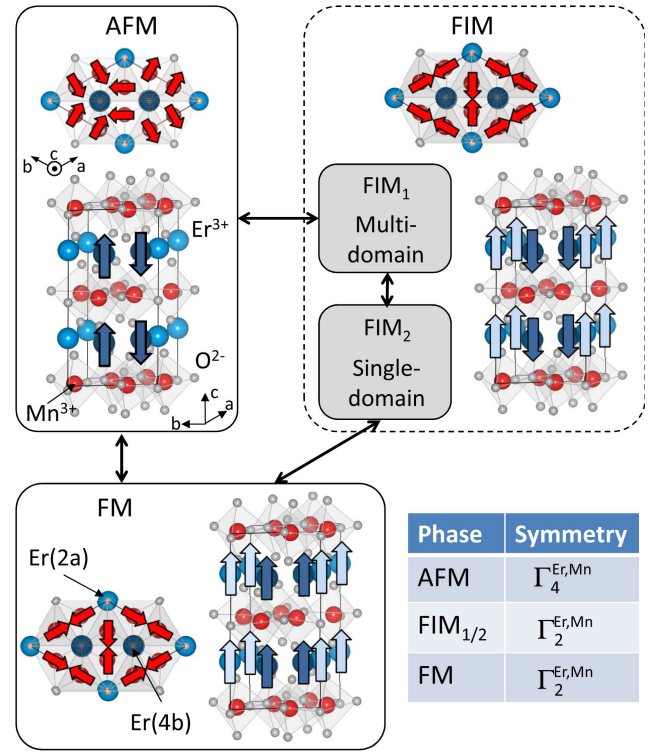}
\caption{\label{fig:fig4} Schematic illustration of the spin structure in the different
magnetically ordered phases of h-ErMnO$_3$. Three phases have to be distinguished: The
antiferromagnetic (AFM) phase, the ferrimagnetic (FIM) phase (magnetic ground state), and the
magnetic-field-induced ferromagnetic (FM) phase. Note that depending on the strength of the
magnetic-field $H||c$ the crystal can either be in a single-domain (FIM$_2$) or a multi-domain
(FIM$_1$) state in the FIM phase. Black arrows between the different phases indicate possible
temperature / magnetic-field driven transitions.}
\end{figure}

Supplementary simulations by SIMREF2.6, however, reveal that for Mn$^{3+}$ order according to
$\Gamma_{2,4}^{\rm{Mn}}$ the experimentally observed (100) reflection only arises if the Er$^{3+}$
sublattice orders according to the $\Gamma_2^{\rm{Er}}$ or $\Gamma_4^{\rm{Er}}$ representation. In
turn it is forbidden for $\Gamma_1^{\rm{Er}}$ and $\Gamma_3^{\rm{Er}}$ as shown by the simulations
in Fig.~\ref{fig:fig3}(b). We further note that $\Gamma_2^{\rm{Er}}$ involves ferromagnetic order
on the 2a and 4b sites and is therefore associated to a macroscopic magnetization which can be
excluded for temperatures between 10~K and $T_{\rm N}$ based on the magnetization
measurement in Fig.~\ref{fig:fig3}(c). We therefore conclude ``compatible'' magnetic order of the
Er$^{3+}$ and Mn$^{3+}$ sublattices according to the same representation, $\Gamma_4^{\rm{Er,Mn}}$,
at $T_{\rm N}$ with an antiferromagnetically ordered Er$^{3+}$(4b) site and a disordered
Er$^{3+}$(2a) site as sketched in Fig.~\ref{fig:fig4} (AFM-phase). This is qualitatively the same
scenario met in h-YbMnO$_3$,\cite{Fabreges08a} but different from
h-DyMnO$_3$,\cite{Wehrenfennig10a} and h-HoMnO$_3$.\cite{Lottermoser04a,Vajk05a}

Below 10~K the pronounced increase of the (100) peak intensity indicates a change in the magnetic
Er$^{3+}$ order. This change is matched by the increase in $M(T)$ seen in Fig.~\ref{fig:fig3}(c).
The only possible transition that is in accordance with the neutron-diffraction and the
magnetization measurements is a $\Gamma_4^{\rm{Er}}\to\Gamma_2^{\rm{Er}}$ transition of the
Er$^{3+}$ sublattices. First of all, as mentioned before, the (100) reflection exclusively occurs
for the representations $\Gamma_2^{\rm{Er}}$ and $\Gamma_4^{\rm{Er}}$. Regarding the two remaining
representations only $\Gamma_2^{\rm{Er}}$ allows for the magnetization observed at low
temperature. This narrows the number of possible magnetic structures down to two, having either an
antiparallel ($\Gamma_2^{\rm{Er}}(\rm{FIM})$) or parallel ($\Gamma_2^{\rm{Er}}(\rm{FM})$)
alignment of the Er$^{3+}$ moments on the 4b and 2a sites. The SIMREF2.6 simulations summarized in
Fig.~\ref{fig:fig3}(b), however, clearly discard Er$^{3+}$ order according to
$\Gamma_2^{\rm{Er}}(\rm{FM})$ because in this case destructive interference of the Er$^{3+}$(4b)
and Er$^{3+}$(2a) contributions suppress the (100) reflection. In contrast, constructive
interference occurs for an antiparallel orientation of the Er$^{3+}$(4b) and Er$^{3+}$(2a) spins
and explains the drastic increase of the (100) peak intensity seen in Fig.~\ref{fig:fig3}(a). We
thus conclude that the ground state of the Er$^{3+}$ order is ferrimagnetic,
$\Gamma_2^{\rm{Er}}(\rm{FIM})$, with antiparallel orientation of the Er$^{3+}$ moments on the 4b
sites relative to the 2a sites (FIM-phase in Fig.~\ref{fig:fig4}). In terms of representations,
the magnetic order of h-ErMnO$_3$ at low temperature is again the same as in
h-YbMnO$_3$\cite{Fabreges08a} and different from h-DyMnO$_3$,\cite{Wehrenfennig10a} and
h-HoMnO$_3$.\cite{Lottermoser04a,Vajk05a}

\subsection{Microscopic magnetic structure of the Er$^{3+}$ and Mn$^{3+}$ sublattices} \label{structure}

Based on the magnetization, SHG, and neutron-diffraction experiments we can now derive a coherent
model describing the different magnetic states indicated in the phase diagram in
Fig.~\ref{fig:fig1}(c). We have already seen that the Mn$^{3+}$ moments and the Er$^{3+}$ moments
on the 4b sites order antiferromagnetically in the AFM-phase while the Er$^{3+}$(2a) spins remain
disordered. The order corresponds to the $\Gamma_4^{\rm{Er,Mn}}$ representation and is
schematically depicted in Fig.~\ref{fig:fig4}. The FIM$_1$ phase encountered toward 0~K in zero
magnetic field is characterized by antiferromagnetic Mn$^{3+}$ order and ferrimagnetic Er$^{3+}$
order, both according to the $\Gamma_2^{\rm{Er,Mn}}$ representation.

Additional information on the nature of this phase can be extracted from the $M(H)$ measurement
presented in Fig.~\ref{fig:fig1}(b). The pronounced response to small magnetic fields ($H\lesssim
0.03$~T) points to the formation of a ferrimagnetic multi-domain state in zero magnetic field with
an $1:1$ ratio of domains with $+M_z$ and $-M_z$. Consequently, the FIM$_2$ phase denotes the
ferrimagnetic single-domain state with a coercive field of $\approx$0.03~T toward 0~K as boundary
between the FIM$_1$ and FIM$_2$ states. The ferrimagnetic nature of this state is further
reflected by the change in signal occurring at the FIM$_2\to$~FM transition in
Fig.~\ref{fig:fig1}(b). By linearly extrapolating the $M(H)$ data gained within the two phases and
comparing the corresponding $M(0)$ values we find that the magnetization almost triples across the
transition. This behavior can be understood in terms of a change from an antiparallel to a
parallel arrangement of the Er$^{3+}$(4b) spins with respect to the Er$^{3+}$(2a) spins, i.e., a
change of the magnetic moment per unit cell from (4-2)$\cdot\mu_{\rm Er^{3+}}$ to
(4+2)$\cdot\mu_{\rm Er^{3+}}$. The transition to the ferromagnetic Er$^{3+}$ order does not
involve a change of magnetic symmetry and is therefore still described by the
$\Gamma_2^{\rm{Er,Mn}}$ representation.

Note that earlier studies did not distinguish between the FIM$_1$ and FIM$_2$ region of the
ferrimagnetic phase. In contrast, the ferrimagnetic rare-earth order was believed to be suppressed
by small magnetic-fields of only 0.05~T which would misleadingly imply weak exchange between the
Er$^{3+}$ spins.\cite{Yen07a}

After the unique determination of the microscopic magnetic structure of h-ErMnO$_3$ we can now
turn to the intermediary states encountered during the transitions between the AFM, FIM, and FM
phases. These transitions will allow us to draw further conclusions about the coupling between the
$4f$ moments at the Er$^{3+}$(2a) and Er$^{3+}$(4b) sites and the $3d$ moments of Mn$^{3+}$.

\subsection{Magnetic interaction of the Er$^{3+}$ and Mn$^{3+}$ sublattices} \label{transitions}

The comparison of Figs.~\ref{fig:fig2} and~\ref{fig:fig3}(a) already revealed an interaction
between the Mn$^{3+}$ and Er$^{3+}$(4b) sublattices: The former triggers the order in the latter
at $T_{\rm N}$. Below 10~K the Er$^{3+}$(2a) order supplements the Er$^{3+}$(4b) order and
additional transitions occur as seen in the phase diagram in Fig.~\ref{fig:fig1}(c).
Figure~\ref{fig:fig3}(a) reveals that, unlike the Er$^{3+}$(4b) order, the Er$^{3+}$(2a) is not
triggered by the Mn$^{3+}$ order. The Er$^{3+}$ spins at the 2a sites begin to order at higher
temperature ($\sim 10$~K) compared to the observed reorientation of the Mn$^{3+}$ spins ($\sim
2$~K). This precludes the triggering mechanism as it would require identical reorientation
temperatures for the two sublattices.

Hence, we propose that here the Er$^{3+}$(2a) order drives the reordering of the Mn$^{3+}$
sublattice through non-biquadratic triggered (and thus linear) order-parameter coupling. This
assumption is supported by three observations: (i) The magnetic rare-earth order continues to
strengthen toward lower temperature whereas the Mn$^{3+}$ order is already saturated. Therefore
the Mn$^{3+}$ order cannot be responsible for the Er$^{3+}$(2a) order. (ii) The reordering of the
Mn$^{3+}$ sublattice ``follows'' the Er$^{3+}$(2a) order at lower temperature. Therefore the
latter order, which is continuously strengthening with decreasing temperature, must be guiding the
reorientation. (iii) The Mn$^{3+}$ sublattice adopts the emerging $\Gamma_2$-like order of the
Er$^{3+}$(2a) sublattice by undergoing a $\Gamma_4^{\rm{Mn}}\to\Gamma_2^{\rm{Mn}}$ transition.

\begin{figure}[h]
\centering
\includegraphics[width=8.5cm,keepaspectratio,clip]{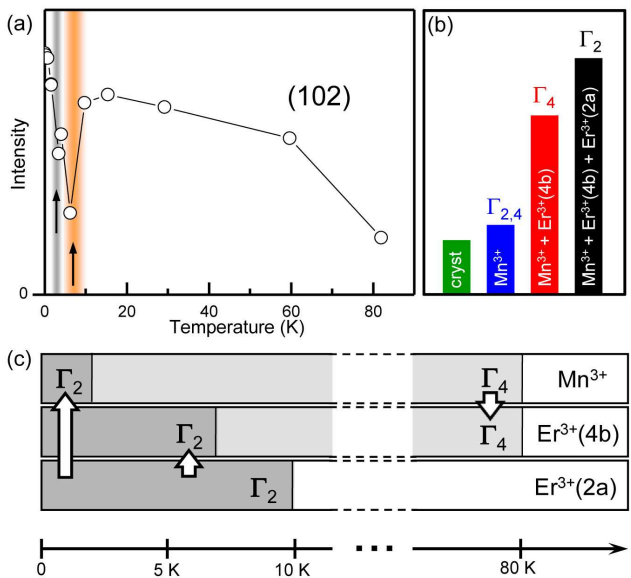}
\caption{\label{fig:fig5} (a) Temperature dependence of the (102) reflection. An increase in
intensity occurs below $T_{\rm N}=80$~K with consecutive breakdowns below 10~K. (b) Decomposition
of contributions to the (102) peak intensity based on SIMREF2.6 simulations. Below $T_N$ the
crystallographic contribution (cryst) to the (102) reflection is supplemented by magnetic
contributions originating from Mn$^{3+}$ and Er$^{3+}$. (c) Sketch illustrating the mutual
induction of the magnetic $3d$ and $4f$ order. Below $T_{\rm N}$ the Mn$^{3+}$ order triggers
magnetic long range order on the 4b sites as indicated by the white arrow. In contrast, magnetic
moments on the 2a sites only order below 10~K inducing a reorientation of Er$^{3+}$(4b) moments
which is followed by a reorientation of the Mn$^{3+}$ spins.}
\end{figure}

The data shown so far do not reveal the role of the Er$^{3+}$(4b) in this interaction: Once the 2a
site orders it dominates the (100) reflection in Fig.~\ref{fig:fig3}(a) to an extent that the
response from the Er$^{3+}$(4b) site is obscured. We therefore investigated various $h0l$
diffraction peaks by SIMREF2.6 simulation in order to identify other magnetic peaks with a similar
weight of contributions from the Mn$^{3+}$, the Er$^{3+}$(2a), and the Er$^{3+}$(2a) that would
allow us to analyzing the interplay between Er$^{3+}$ moments on the 2a and 4b sites. As revealed
by Fig.~\ref{fig:fig5}(b), the (102) reflection fulfills this condition. We find that the
intensity of the (102) reflection experiences striking breakdowns around 2.5~K and 7~K revealing
magnetic reorientations with a transient breakdown of the magnetic order at these
temperatures.\cite{Lonkai04a,Helton11a} The breakdown at 2.5~K is related to the Mn$^{3+}$
sublattice because it is also observed in the (101) and (103) reflexes (see Fig.~\ref{fig:fig2})
which are entirely determined by the Mn$^{3+}$ order. In turn, the additional breakdown at 7~K can
only indicate a magnetic reorientation of the Er$^{3+}$(4b) sublattice. We conclude that, like the
Mn$^{3+}$ order, the Er$^{3+}$(4b) order toward 0~K is driven by the Er$^{3+}$(2a) order because
it follows the same criteria (i-iii) as the Mn$^{3+}$ reordering.

In summary, toward 0~K h-ErMnO$_3$ undergoes a transformation with a change of the representation
according to $\Gamma_4^{\rm{Er,Mn}}\to\Gamma_2^{\rm{Er,Mn}}$. However, the $\Gamma_2$ phase is
assumed at a different temperature for the Er$^{3+}$(2a), the Er$^{3+}$(4b), and the Mn$^{3+}$
sublattices.

\section{Comprehensive model for the magnetic phase transitions of hexagonal E\symbol{114}M\symbol{110}O$_3$} \label{model}

By combining the analysis of the magnetic phase diagram in Sections~\ref{magnetization} to
\ref{structure} and the analysis of the magnetic transitions between these phases in
Section~\ref{transitions} we are now able to present a comprehensive scenario of the magnetic
interactions and the resulting phase transitions in h-ErMnO$_3$ with a projection on the
h-$R$MnO$_3$ series as a whole. We distinguish two fundamentally different temperature ranges as
illustrated in Fig.~\ref{fig:fig5}(c):

\textit{Between the N\'{e}el temperature and $\sim 10$~K:} In this range the magnetic structure is
determined by the ordering of the Mn$^{3+}$ sublattice. The Mn$^{3+}$ spins order
antiferromagnetically at $T_{\rm N}$ and promote rare-earth ordering on the Er$^{3+}$(4b) site.
The Er$^{3+}$(4b) order is induced at the same temperature, $T_{\rm N}$, via a triggering
mechanism with biquadratic Mn$^{3+}$--Er$^{3+}$(4b) order-parameter coupling. Both the Mn$^{3+}$
and Er$^{3+}$(4b) lattice order according to the same representation, $\Gamma_4^{\rm{Er,Mn}}$.
However, because of the biquadratic coupling this compatibility is not mandatory. Indeed,
Mn$^{3+}$ and $R^{3+}$(4b) order according to different representations in h-DyMnO$_3$. In any
case, the Er$^{3+}$(2a) site remains disordered in this temperature range.

\textit{Below $\sim 10$~K:} In this range the magnetic structure is determined by the ordering of
the Er$^{3+}$(2a) sublattice. The spins arrange uniformly at $\sim 10$~K according to the
representation $\Gamma_2^{\rm{Er}}$ with a macroscopic magnetization per domain and a nonzero net
magnetization once the magnetic field lifts the degeneracy between oppositely oriented FIM domain
states. When the Er$^{3+}$(2a) order strengthens toward low temperature, the Er$^{3+}$(2a)
ordering at first induces $\Gamma_4^{\rm{Er}}\to\Gamma_2^{\rm{Er}}$ reordering on the
Er$^{3+}$(4b) site ($\sim 7$~K) followed by $\Gamma_4^{\rm{Mn}}\to\Gamma_2^{\rm{Mn}}$ reordering
of the Mn$^{3+}$ sublattice ($\sim 2.5$~K). The coupling to the Er$^{3+}$(2a) sublattice is linear
and therefore not guided by the triggering mechanism. This is evidenced by the difference between
the (re-) ordering temperatures and the accordance of the representations describing the magnetic
order in the different sublattices. The Er$^{3+}$(2a) and Er$^{3+}$(4b) sites maintain an
antiparallel spin orientation that can be overcome in an external magnetic field in the order of
magnitude of 1~T driving a transition from ferrimagnetic to ferromagnetic Er$^{3+}$ order. A
similar situation is met in all rare-earth h-$R$MnO$_3$ compounds except in h-HoMnO$_3$ where the
Ho$^{3+}$ ground state is antiferromagnetic, yet with similar sublattice correlations as in the
other h-$R$MnO$_3$ compounds.

\section{Summary}

In conclusion, the combination of magnetization, SHG, and neutron-diffraction experiments at
temperatures down to the milli-Kelvin regime reveals important features in the magnetic phase
diagram of h-ErMnO$_3$. We identify the spin structure in the respective magnetic phases, the
magnetic coupling between the Mn$^{3+}$, Er$^{3+}$(4b), and Er$^{3+}$(2a) sublattices, and the
resulting phase transitions establishing the phase diagram. We find a high-temperature range above
10~K where the Mn$^{3+}$ sublattice induces magnetic Er$^{3+}$ order and a low-temperature range
below 10~K where the Er$^{3+}$ sublattice induces magnetic Mn$^{3+}$ order. Most of all, we find
that the ordering on the 4b and 2a sites of the Er$^{3+}$ sublattices play a strikingly
independent and different role in establishing the magnetic order in h-ErMnO$_3$. The
comprehensive model for the phase diagram of the h-ErMnO$_3$ developed in this work can be
projected onto the other rare-earth h-$R$MnO$_3$ compounds with only small variations. Thus, we
are now able to understand the complex magnetic phases of the h-$R$MnO$_3$ system on a universal
basis.

D.M. acknowledges support by the Alexander von Humboldt Foundation and the NSF Science and Technology Center (E3S). M.F. thanks the DFG (SFB 608) for subsidy.



\begin{thebibliography}{}
\bibitem{footnote1}
Based on Section~\ref{Er-order} the SIMREF2.6 simulation was performed assuming Er$^{3+}$ order according to the $\Gamma_2^{\rm{Er}}$ and $\Gamma_4^{\rm{Er}}$ representation.
\end{thebibliography}
\end{document}